\documentclass[runningheads]{llncs}
\usepackage{type1cm}        
\usepackage{makeidx}         
\usepackage{graphicx}        
\usepackage{multicol}        
\usepackage[bottom]{footmisc}

\usepackage{newtxtext}       %
\usepackage{newtxmath}       

\usepackage{amsmath}
\usepackage{algorithm}
\usepackage{algpseudocode}
\usepackage{subfig}
\usepackage{graphicx}

\usepackage[version=4]{mhchem}
\usepackage{siunitx}
\usepackage{longtable,tabularx}
\usepackage{subfiles} 
\usepackage{pdfpages} 
\usepackage{csvsimple} 
\usepackage{tikz} 
\usetikzlibrary{shapes, arrows, positioning} 
\usepackage{hyperref}

\usepackage[T1]{fontenc}
\usepackage{graphicx}
\begin{document}
\title{Fuzzy Decisions on Fluid Instabilities: Autoencoder-Based Reconstruction meets Rule-Based Anomaly Classification}

\titlerunning{AER and Fuzzy for fluid disruption data}
%
\author{Bharadwaj Dogga\inst{1}\orcidID{0009-0002-5915-1367} \and
Gibin Raju\inst{2}\orcidID{0000-0003-2559-6931} \and
Wilhelm Louw\inst{3}\orcidID{0009-0009-8683-1932} \and
Kelly Cohen\inst{4}\orcidID{0000-0002-8655-1465}}
\authorrunning{B. Dogga et al.}
%
\institute{Ph.D. Student, AI Bio Lab, Digital Futures, University of Cincinnati, \\Cincinnati, OH 45221, USA e-mail:\email{doggabj@mail.uc.edu} \and
Postdoctoral Researcher, Department of Multidisciplinary Engineering, TA\&M University, \\College Station, TX 77843, USA e-mail: \email{gibinraju@tamu.edu} \and
Research Associate AI Bio Lab, Digital Futures, University of Cincinnati, \\Cincinnati, OH 45221, USA e-mail:\email{louwwa@ucmail.uc.edu} \and
Director AI Bio Lab, Digital Futures, University of Cincinnati, \\Cincinnati, OH 45221, USA e-mail:\email{cohenky@ucmail.uc.edu}
}
\maketitle              
\begin{abstract}

Shockwave classification in shadowgraph imaging is challenging due to limited labeled data and complex flow structures. This study presents a hybrid framework that combines unsupervised autoencoder models with a fuzzy inference system to generate and interpret anomaly maps. Among the evaluated methods, the hybrid $\beta$-VAE autoencoder with a fuzzy rule-based system most effectively captured coherent shock features, integrating spatial context to enhance anomaly classification. The resulting approach enables interpretable, unsupervised classification of flow disruptions and lays the groundwork for real-time, physics-informed diagnostics in experimental and industrial fluid applications.

\end{abstract}

\section{Introduction and Related Work}
\label{sec:1}

Image classification in shadowgraph and schlieren imaging plays a critical role in studying unsteady compressible flows, boundary layer interactions, and dynamic fluid phenomena, particularly in high-speed aerodynamics and industrial diagnostics \cite{Maharjan2021-oz}. Traditional image processing approaches often rely on edge detection techniques—such as Canny, Isolation Forests, or Sobel—to extract shock features from experimental visualizations \cite{Zheng2021-mf}\cite{Li2021-os}\cite{Znamenskaya2020-pg}. While effective in idealized conditions, these methods require careful pre-processing, are sensitive to threshold selection, and often degrade under noisy or complex shock geometries.
To address these issues, recent studies have explored deep learning methods that improve robustness and generalizability. Convolutional neural networks (CNNs), for example, have been trained to directly identify shock regions from raw shadowgraph images, showing improved performance in unstructured settings \cite{Liu2024-zv}. In parallel, unsupervised learning approaches using autoencoders, especially variational autoencoders (VAEs), have shown promise for anomaly detection by modeling data distributions and flagging deviations through reconstruction errors—effective across domains such as industrial inspection and medical imaging \cite{Baur2021-yg}.
Despite these advancements, deep models offer limited support for classifying anomaly type or severity which is critical for physics-based systems where decision making relies on explainable outputs. Hybrid models combining neural networks with fuzzy logic have been proposed to bridge this gap, particularly in video surveillance and uncertainty aware anomaly detection \cite{Muhammad2022-hk}. These systems offer human-like decision rules but are rarely adapted for physical classification tasks.
To overcome these limitations, this study proposes a hybrid framework that integrates autoencoder variants with a fuzzy inference system to enhance anomaly classification in fluid shadowgraph images. The goal is to enable unsupervised, interpretable classification of shock-related structures in fluid diagnostics. This approach contributes a novel application of fuzzy reasoning to autoencoder error interpretation in compressible flows, with targeted rule design that enables unsupervised classification of shock probability without the need for ground-truth labels.

\section{Methodology}
\label{sec:2}

The proposed hybrid framework integrates an image-based autoencoder with a Mamdani-style Fuzzy Inference System (FIS) to enable unsupervised detection and classification of shock anomalies in shadowgraph images. The autoencoders reconstruct the input image, while the FIS evaluates reconstruction and spatial errors to classify detected anomalies into physically interpretable shock probabilities.

\subsection{Autoencoder Architectures and Image Quality Metrics}
\label{subsec:2}

Two unsupervised autoencoder models were evaluated in this study: a Denoising Autoencoder (DAE) and a $\beta$-Variational Autoencoder ($\beta$-VAE). Both models were trained to reconstruct grayscale shadowgraph images with the goal of identifying flow anomalies through reconstruction errors. To assess performance, three image quality metrics were computed: Mean Squared Error (MSE), Structural Similarity Index Measure (SSIM), and Shannon entropy of the error maps. MSE quantifies the average pixel-wise difference between the original and reconstructed images, while SSIM accounts for luminance, contrast, and structure, providing a perceptual measure of similarity. Entropy captures the spatial complexity of the reconstruction error, serving as a proxy for anomaly dispersion. Together, these metrics offer quantitative insights into model fidelity and anomaly localization.

\subsection{Error Map Computation and Fuzzy Rule-Based System}

Although the reconstruction metrics described earlier provide a quantitative basis for comparing model performance, additional parameters are required to spatially resolve and classify anomalies in shadowgraph images. These parameters are defined below:

\begin{enumerate}
        \item Pixel error: This is defined as the absolute error value between reconstructed image and input image normalized by max value at the pixel
        \item Neighborhood error: The above pixel error values are first normalized using min-max normalization technique and then spatial averaging is done with a uniform filter of $kernel\_size=5$.
\end{enumerate}
\vspace{-1em} 

\begin{algorithm}[H]
\caption{Fuzzy Inference System for Anomaly Classification}
\label{alg1}
\begin{algorithmic}[1]

\State \textbf{Define Input Variables and Membership Functions:} 

\textit{Pixel Error} ($\texttt{pixel\_err}$) with domain $[0, 1]$:
\begin{align*}
\mu_{\text{low}}(x) &= \text{trimf}(x; 0, 0, 0.3) \\
\mu_{\text{medium}}(x) &= \text{trimf}(x; 0.2, 0.5, 0.8) \\
\mu_{\text{high}}(x) &= \text{trimf}(x; 0.7, 1, 1)
\end{align*}

\textit{Neighborhood Error} ($\texttt{neigh\_err}$) with domain $[0, 1]$:
\begin{align*}
\mu_{\text{low}}(x) &= \text{trimf}(x; 0, 0, 0.3) \\
\mu_{\text{medium}}(x) &= \text{trimf}(x; 0.2, 0.5, 0.8) \\
\mu_{\text{high}}(x) &= \text{trimf}(x; 0.7, 1, 1)
\end{align*}

\State \textbf{Define Output Variable and Membership Functions:}

\textit{Anomaly} ($\texttt{anomaly}$) with domain $[0, 1]$:
\begin{align*}
\mu_{\text{none}}(x) &= \text{trimf}(x; 0, 0, 0.3) \\
\mu_{\text{strong}}(x) &= \text{trimf}(x; 0.2, 0.5, 0.8) \\
\mu_{\text{possible}}(x) &= \text{trimf}(x; 0.7, 1, 1)
\end{align*}

\State \textbf{Fuzzy Rules:}

\begin{enumerate}
    \item If $\texttt{pixel\_err}$ is \texttt{high} and $\texttt{neigh\_err}$ is \texttt{high}, then $\texttt{anomaly}$ is \texttt{strong}.
    \item If $\texttt{pixel\_err}$ is \texttt{medium} and $\texttt{neigh\_err}$ is \texttt{medium}, then $\texttt{anomaly}$ is \texttt{possible}.
    \item If $\texttt{pixel\_err}$ is \texttt{high} and $\texttt{neigh\_err}$ is \texttt{low}, then $\texttt{anomaly}$ is \texttt{possible}.
    \item If $\texttt{pixel\_err}$ is \texttt{low} and $\texttt{neigh\_err}$ is \texttt{high}, then $\texttt{anomaly}$ is \texttt{possible}.
    \item If $\texttt{pixel\_err}$ is \texttt{low} and $\texttt{neigh\_err}$ is \texttt{low}, then $\texttt{anomaly}$ is \texttt{none}.
\end{enumerate}

\end{algorithmic}
\end{algorithm}

Using both errors types, a Mamdani-type fuzzy system is implemented using triangular membership functions, as outlined in Algorithm \ref{alg1}. In this formulation, high anomaly likelihood is assigned to intermediate error values (e.g., 0.2–0.8), reflecting regions of structured deviation rather than extreme noise or perfect reconstruction.

\section{Experimental Setup}
\label{subsec:3}

Experiments were conducted at the University of Cincinnati's Heated Jet Noise Facility used 15,000 frames across 15 flow conditions\cite{raju_g}. Dry air, regulated from storage tanks, was supplied to circular nozzle with a nozzle pressure ratio of 3.6 and temperature ratios of 1.0. High-speed shadowgraph imaging employed a Z-type setup with 12-inch parabolic mirrors, ORIEL UV light sources, and Phantom $v1610/v1210$ cameras, capturing 1000 frames per run at 25,000 fps with $1.5$-$2.0 \mu s$ exposure. A random image section was then imported as a test bed for the framework.

\begin{figure}[hbt!]
\centering
\includegraphics[width = 0.14\linewidth]{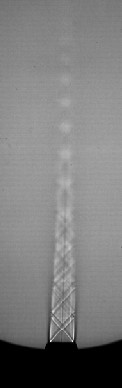}
\caption{Test section image from Shadowgraph with various shock phenomenon}
\label{fig:original_image}
\end{figure}

\section{Results and Discussion}
\label{sec:3}

This section presents both quantitative and qualitative evaluation of shock anomaly classification using classical baseline methods and the proposed autoencoder–fuzzy hybrid framework described in Section~\ref{sec:2}.

To quantitatively benchmark autoencoder-based methods, we evaluate three conventional approaches: Canny edge detection, Isolation Forest, and Sobel gradient magnitude. As shown in Table \ref{tab:quantative_metrics}, all baselines yield higher MSE and lower SSIM than the deep learning models. Canny and Isolation Forest are prone to oversensitivity and lack global context, while Sobel offers smoother gradients but limited semantic awareness.Their lack of continuity and contextual reasoning underscores the need for a more robust hybrid approach using autoencoders and fuzzy logic. 

\subsection{Autoencoder Model Performance and FIS Post-Processing}

Deep learning methods demonstrate significantly better anomaly localization and image structure preservation. The Denoising Autoencoder (DAE) achieves balanced results across all metrics, while the $\beta$-Variational Autoencoder ($\beta$-VAE) records the best SSIM (0.1676) and highest entropy (10.3085), capturing broad structural anomalies relevant to shock regions. DAE, though slightly behind in SSIM, benefits from spatial sharpness in error concentration. These results validate the utility of autoencoders in capturing fluid anomalies with greater fidelity than handcrafted features.

\begin{table}[h!]
\centering

\begin{tabular}{l|c|c|c}
\textbf{Method} & \textbf{MSE} & \textbf{SSIM} & \textbf{Entropy} \\
\hline
&&\\
Canny Edge Detection    & 0.2722 & 0.0833 & 6.7004 \\
Isolation Forest & 0.2830 & 0.0833 & 7.7785 \\
Sobel Gradient Magnitude     & 0.2499 & 0.1332 & 9.9844 \\
Denoising AE       & 0.2579 & 0.0791 & 10.0010 \\
$\beta$-Variational AE  & 0.2241 & 0.1676 & 10.3085

\end{tabular}
\vspace{0.5em}
\caption{Quantitative Evaluation of Anomaly Detection Methods}
\label{tab:quantative_metrics}
\end{table}

\begin{figure}[hbt!]
\centering
\includegraphics[trim=0 0mm 00 10mm,clip,width = 0.75\linewidth]{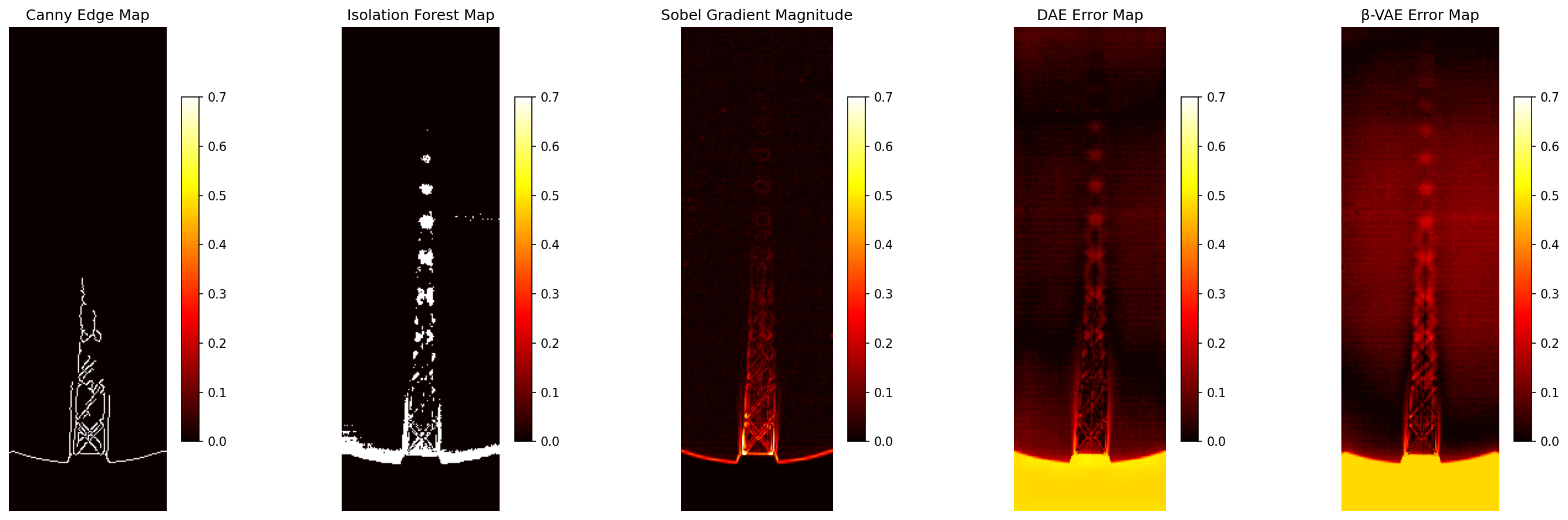}
\caption{Error map, from left to right: Canny, Isolation Forest, Sobel, DAE, $\beta-$VAE}.
\label{fig:aer_error_map}
\end{figure}

Figure \ref{fig:aer_error_map} qualitatively compares five methods for shock detection via error mapping. Classical methods such as Canny and Isolation Forest produce either fragmented or over-saturated outputs, while Sobel offers smoother gradients but lacks semantic filtering. In contrast, DAE localizes sharp, narrow anomalies aligned with shocks, whereas $\beta$-VAE captures broader, more coherent structures. These autoencoder maps provide a more reliable basis for downstream classification.

\begin{figure}[hbt!]
\centering
\includegraphics[trim=0 0mm 00 30mm,clip,width = 0.45\linewidth]{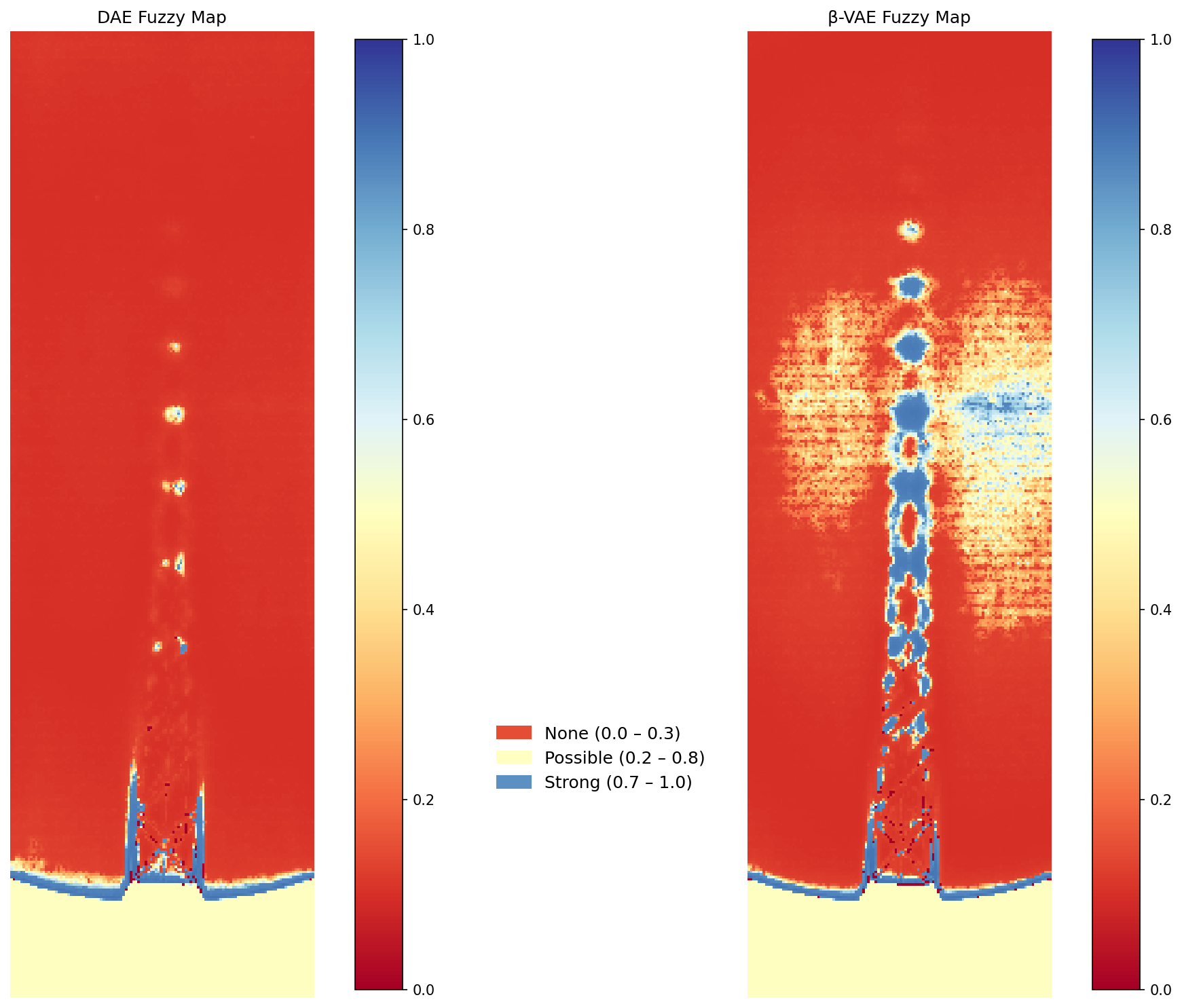}
\caption{Fuzzy rule-based shock anomaly map, from left to right: DAE, $\beta-$VAE.}
\label{fig:fuzzy_rule_error_map}
\end{figure}

To leverage these enhanced error representations for region-based classification, a Fuzzy Inference System (FIS) was applied to the AE error maps. These results are shown in Figure \ref{fig:fuzzy_rule_error_map}. The fuzzy map from DAE clearly isolates sharp, discrete anomalies that match the high-curvature points in the flow field. The $\beta$-VAE fuzzy map, however, extends smoothly across the full shock column with high anomaly confidence ($>0.7$), classifying the shock zones with greater physical consistency.

The fuzzy-enhanced anomaly maps reveal distinct model behavior. DAE identifies sharp, localized disruptions aligned with individual shocks but is prone to false positives from background noise. In contrast, $\beta$-VAE offers smoother, more coherent anomaly fields that align closely with the full shock structure. The fuzzy inference system effectively consolidates spatial and reconstruction errors, suppressing noise and highlighting dominant features without requiring labeled data. This makes the $\beta$-VAE hybrid approach well-suited for unsupervised anomaly classification in fluid flow diagnostics.

Future work will focus on improving computational efficiency and reducing time cost. The fuzzy system can be further optimized by fine-tuning membership functions or adopting Sugeno inference to enable real-time execution. Transformer-based architectures may offer more robust feature extraction for complex flow fields. Additionally, incorporating turbulence modeling and fluid conditions to enhance FIS parameter tuning will improve classification accuracy for complex shock-turbulence interactions.

\section{Conclusion}
\label{sec:5}

This study proposes a hybrid framework that combines unsupervised autoencoders with fuzzy inference systems for anomaly classification in fluid shadowgraph images. Among the evaluated models, $\beta$-VAE demonstrated the best performance in capturing coherent shock structures, when compared with standard baseline methods and Denoising AE variant. The fuzzy rule-based system improved interpretability by incorporating spatial context into the anomaly scoring process. Together, these components enable unsupervised and physically meaningful anomaly classification without the need for labeled data, making the approach well-suited for experimental fluid diagnostics. This framework lays the groundwork for real-time, physics-informed analysis of complex flow phenomena.

\begin{credits}

\subsubsection{\discintname}
The authors have no competing interests to declare that are
relevant to the content of this article. 
\end{credits}
%
%
%

\begin{thebibliography}{8}
\bibitem{Zheng2021-mf}Zheng, L., Lawlor, B., Katko, B., McGuire, C., Zanteson, J. \& Eliasson, V. Image processing and edge detection techniques to quantify shock wave dynamics experiments. {\em Exp. Tech.}. \textbf{45}, 483-495 (2021,8)

\bibitem{Maharjan2021-oz}Maharjan, S., Bjerketvedt, D. \& Lysaker, O. Processing of high-speed videos of shock wave boundary layer interactions. {\em Signal Image Video Process.}. \textbf{15}, 607-615 (2021,4)

\bibitem{Muhammad2022-hk}Muhammad, K., Obaidat, M., Hussain, T., Ser, J., Kumar, N., Tanveer, M. \& Doctor, F. Fuzzy logic in surveillance Big Video Data analysis. {\em ACM Comput. Surv.}. \textbf{54}, 1-33 (2022,4)

\bibitem{Baur2021-yg}Baur, C., Denner, S., Wiestler, B., Navab, N. \& Albarqouni, S. Autoencoders for unsupervised anomaly segmentation in brain MR images: A comparative study. {\em Med. Image Anal.}. \textbf{69}, 101952 (2021,4)

\bibitem{Liu2024-zv}Liu, J., Xie, G., Wang, J., Li, S., Wang, C., Zheng, F. \& Jin, Y. Deep industrial image anomaly detection: A survey. {\em Mach. Intell. Res.}. \textbf{21}, 104-135 (2024,2)

\bibitem{Znamenskaya2020-pg}Znamenskaya, I., Doroshchenko, I. \& Tatarenkova, D. Edge detection and machine learning approach to identify flow structures on schlieren and shadowgraph images. {\em Proceedings Of The 30th International Conference On Computer Graphics And Machine Vision (GraphiCon 2020). Part 2}. pp. aper15-1–paper15–14 (2020,12)

\bibitem{Li2021-os}Li, G., Kontis, K. \& Fan, Z. Automatic shock detection, extraction, and fitting in schlieren and shadowgraph visualization. {\em AIAA J.}. \textbf{59}, 2312-2317 (2021,6)

\bibitem{raju_g}Raju, G., Baier, F., Mora, P. \& Gutmark, E. Shock-cell spacing analysis in heated jets using shadowgraph technique. {\em Conference: International Conference on Jets, Wakes and Separated Flows}. (2017,10)

\end{thebibliography}
%

\end{document}